\documentclass[11pt,journal,compsoc,twocolumn]{IEEEtran}
\usepackage[]{hyperref}
\usepackage[square]{natbib}
\renewcommand{\cite}{\citep}

\usepackage{graphicx}
\usepackage[all,british]{foreign}
\usepackage{multirow, multicol, bigstrut}
\usepackage{longtable}
\usepackage{tabularx}
\usepackage{comment}
\usepackage{color}
\usepackage[warn]{textcomp}

\newcommand{\hl}{}

\begin{document}
%
\title{Prevalence of Potentially Predatory Publishing\\ in Scopus on the Country Level}



\author{    
    \IEEEauthorblockN{
        Tatiana~Savina\IEEEauthorrefmark{2}\IEEEauthorrefmark{1}
        \hspace{4cm}
        Ivan~Sterligov\IEEEauthorrefmark{2}
    }
    
    \IEEEauthorblockA{
        \IEEEauthorrefmark{2} National Research University Higher School of Economics, Moscow, Russia
    }
    
    \IEEEauthorblockA{
        \IEEEauthorrefmark{1}Russian Foundation for Basic Research, Moscow, Russia.
    }
    
    \texttt{\small{savina.tf@gmail.com \hspace{4cm} isterligov@hse.ru}}

}
\markboth{Prevalence of Potentially Predatory Publishing in Scopus on the Country Level}%
{}
%


\maketitle

\begin{abstract}
We present results of a large-scale study of \textit{potentially predatory journals} (PPJ) represented in the Scopus database, which is widely used for research evaluation. Both journal metrics and country/disciplinary data have been evaluated for different groups of PPJ: those listed by Jeffrey Beall and those discontinued by Scopus because of ``publication concerns''. Our results show that even after years of discontinuing, hundreds of active potentially predatory journals are still highly visible in the Scopus database. PPJ papers are continuously produced by all major countries, but with different prevalence. Most ASJC (All Science Journal Classification) subject areas are affected. The largest number of PPJ papers are in engineering and medicine. On average, PPJ have much lower citation metrics than other Scopus-indexed journals. We conclude with a  survey of the case of Russia and Kazakhstan where the share of PPJ papers in 2016 amounted to almost a half of all Kazakhstan papers in Scopus. Our data suggest a relation between PPJ prevalence and national research evaluation policies. As such policies become more widespread, the expansion of potentially predatory journal research will be increasingly important.
\end{abstract}

\begin{IEEEkeywords}
potentially predatory journals, government publishing policy
\end{IEEEkeywords}


\section{Introduction}\label{sec:introduction}

\IEEEPARstart{R}{ecent} years have witnessed many dramatic changes in scholarly communication across the world. The main drivers of these changes are the globalization of academia and proliferation of the Internet and digital technologies as well as the spread of the evaluation culture in research management \cite{dahler-larsen2011a}. The publish-or-perish motto \cite{roland2007a, steele2006a} and an ever-increasing supply of available metrics \cite{weingart2005a, wilsdon2015a} have facilitated the rapid growth of ``citizen bibliometrics'' including the usage of scientometric indicators by administrators of various degree of competence as well as by researchers themselves \cite{leydesdorff2016a}. In short, research evaluation has become substantially more formalized relying on various indicators, which are mostly based on publication and citation counts. This metrics explosion is partially attributed to the priorities of many nations and organizations to reproduce the success of world leaders in science and technology.

In this paper, we provide a bird's eye view of the growth of articles in potentially predatory journals (PPJ), a global phenomenon stemming from both the evaluation/metrics culture and globalization. ``\textit{Potentially predatory journals}'' refers to publication venues providing publishing services of questionable quality while having the formal characteristics of respected academic journals, usually to receive article processing fees.

We introduce an algorithm that matches Jeffrey Beall's lists  \cite{beall2016publishers, beall2016journals} of publishers and standalone journals with the Source title list from Scopus and achieve a significantly better rate of detection than in previous studies \cite{bagues2017a, machacek2017a, machacek2019a}. This allows us to show a comprehensive breakdown of such publications by countries and disciplines over time, and provide an overview of indicators for PPJ, as compared to non-PPJ sources. We argue that the PPJ phenomenon is mainly due to proliferation of indicator-based research evaluation systems \cite{glaeser2010a} and is an example of a goal displacement \cite{colwell2012a, rijcke2016a} and opportunistic behavior \cite{oender2017a}. In contrast to the previous work \cite{shen2015a}, our data suggest that almost all countries are affected, although with marked differences in the share of PPJ articles by country.

First, we provide an overview of Scopus usage across the world in order to justify choosing this particular data source. Next, we describe the PPJ phenomenon and review emerging studies of it. We pay special attention to the peculiarities and limitations of Jeffrey Beall's lists. We also describe our data and methods. Then, the main section describes global PPJ statistics in Scopus and also reviews some recent efforts by Elsevier to clean its database. We conclude with an analysis of possible causes of the observed drastic variations in relative PPJ publications across various nations and sum up with policy advice. 

\subsection{Scopus in Research Evaluation}

Before talking about potentially predatory journals, it is important to briefly describe the reasons of Scopus to be attractive for publishers and authors. Scopus is widely used in research evaluation. Along with the \textit{Web of Science} (WoS) and \textit{Google Scholar}, Scopus is a leading resource for searching relevant works and evaluation of researchers and journals \cite{wouters2015a}. These three databases, however, have different indexing policies and marketing strategies, leading to different coverage and indicator values (for recent comparisons see \cite{moed2016a, mongeon2016a, harzing2016a}). Google Scholar is not widely used by research managers, because of its strategy of covering virtually all scientific literature with little quality control. So the options suitable for research evaluation are the Web of Science and Scopus.

The Web of Science is the most exclusive in terms of the number of indexed journals and poses itself as a ``painstakingly selected, actively curated database of the journals that researchers themselves have judged to be the most important and useful in their fields''\footnote{Promotional info at \url{http://clarivate.com/?product=web-of-science} accessed on 20 june 2017}. \hl{Scopus' promotional text is similar, with a special reference of their customers' high academic standards: ``Scopus has a clearly stated selection policy and an internationally acclaimed board of selection experts so you can be sure that what you see on Scopus meets your high standards''}\footnote{Promotional info at \url{https://www.elsevier.com/solutions/scopus/content/content-policy-and-selection} accessed on 21 June 2017}. \hl{Judging from the sheer number of indexed journals, Scopus has more inclusive criteria, which is perceived as intentional and central to its main marketed advantage over WoS, \ie the scope, reflected in its name itself.} \citet{mongeon2016a} find that Scopus has a ``larger journal coverage in all fields'', especially in Business \& Management and in Social Sciences. Country-level data presented by these authors also is favorable for Scopus for all 15 major countries studied both in the number of published journals and in the number of published papers. Thus, we conclude that in recent years Scopus has outperformed WoS in terms of coverage in all or at least in the vast majority of disciplines.

Also, Scopus has been chosen instead of WoS for several high-profile nation-wide evaluation projects, namely Research Excellence Framework in the United Kingdom, Excellence in Research for Australia (until 2019), and Evaluation of R\&D Units in Portugal. As the Australian Research Council (ARC) put it in their press release, ``When selecting Scopus, the ARC had regard to their coverage of relevant journals...''\footnote{See \url{ http://www.arc.gov.au/news-media/media-releases/scopus-provide-citation-information-era} accessed on 21 June 2017 and currently available only via Internet Archive: \url{ https://web.archive.org/web/20151010083040/https://www.arc.gov.au/news-media/media-releases/scopus-provide-citation-information-era}}. Scopus has recently replaced WoS as a bibliometric data source for the influential Times Higher Education World University Rankings, and is already used by QS World University Rankings.

\begin{figure}
\includegraphics[width=\linewidth]{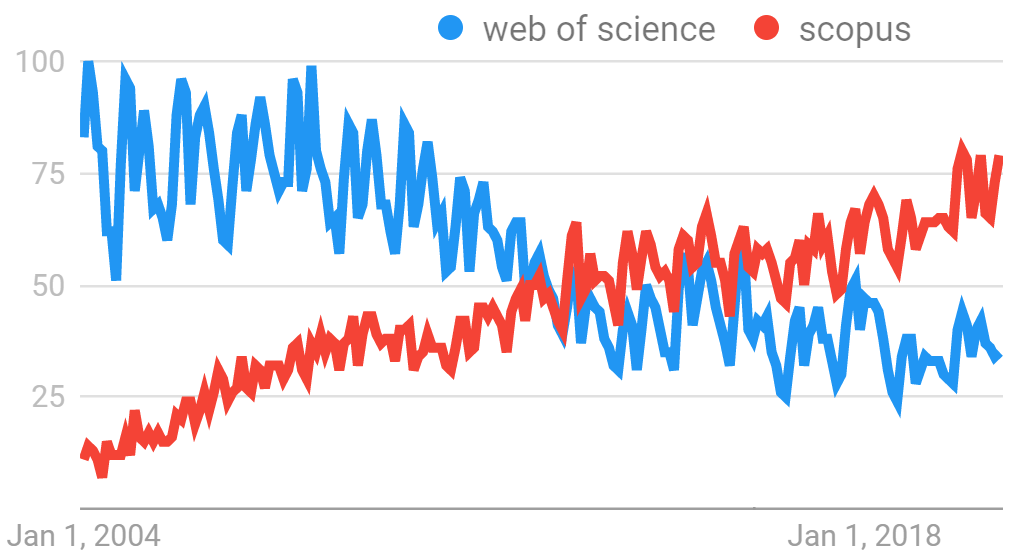}
\caption{\label{fig:google_trends_time} Monthly Google Trends* data for search queries ``Web of Science'' (blue) vs ``Scopus'' (red) in 2004-2019. 
The horizontal axis of the main graph represents time, and the vertical is how often a term is searched for relative to the total number of searches, globally.}

\includegraphics[width=\linewidth]{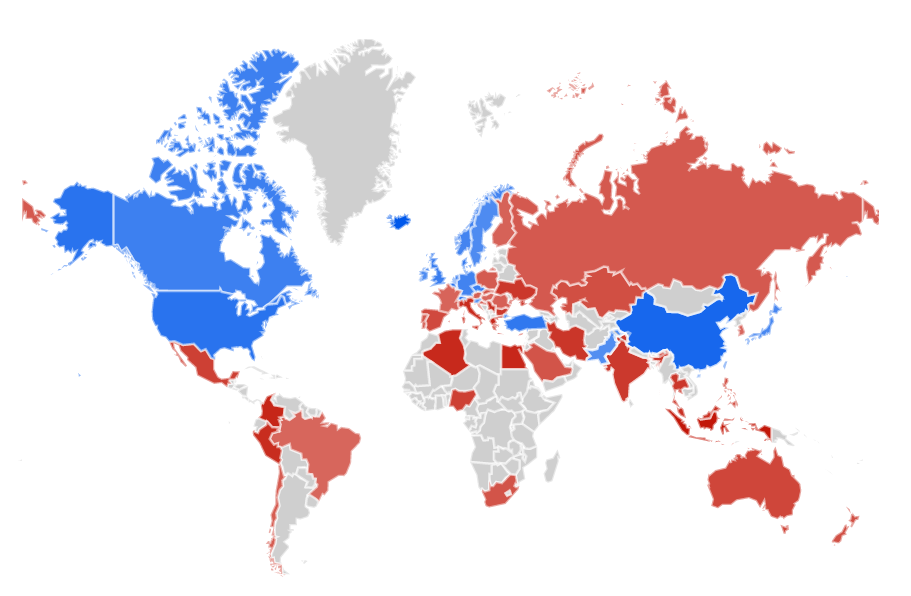}
\caption{\label{fig:google_trends_countries} The popularity of Google Search* queries for Web of Science vs Scopus since 2004. The color intensity represents the percentage of searches. Search term popularity is relative to the total number of Google searches performed at a specific time, in a specific location.\protect\\\protect\\
**\url{https://trends.google.com/trends/explore?date=all&q=web\%20of\%20science,scopus}}
\end{figure}

Overall, Scopus can now be considered as a more popular bibliometrics database, if we take into account Google Trends data that reflects global search intensity (see Figure~\ref{fig:google_trends_time}). This metric is plausible because there are almost no meanings of the terms ``Scopus'' and ``Web of Science'' other than those related to the two databases\footnote{While ``Web of Science'' usually clearly means relevant database, ``Scopus'' may also refer to Mount Scopus, a historical mountain in northeast Jerusalem, or a latin name of the bird species Hamercop (Scopus Umbretta), or a specialist journal on east African ornithology.}.

The country-level data may be more relevant to the topic of our paper (see Figure~\ref{fig:google_trends_countries}), which show that in some regions people search for WoS and Scopus much more than in the others and that Scopus leads the way in many of the countries studied in the results section. Although vague, Google Trends data show that the general interest is substantial.

\subsection{Potentially Predatory Journals}

The PPJ phenomenon and its naming are both directly linked to the work of Jeffrey Beall, an academic librarian and associate professor of library science at the University of Colorado, Denver. In 2010, he published a review of Bentham Science Publishers in ``The Charleston Advisor'' \cite{beall2009a}, which ended with a verdict that is typical of Beall's latter critique of PPJ: ``Bentham Open's emergence into scholarly publishing in 2007 has served mainly as a venue to publish research of questionable quality. The site has exploited the Open Access (OA) model\footnote{Open access scholarly literature is free of charge and often carries less restrictive copyright and licensing barriers than traditionally published works, for both the users and the authors.} for its own financial motives and flooded scholarly communication with a flurry of low quality and questionable research...''.

It is important to note that Beall clearly links the problem to the Gold Open Access (``author pays'') model of journal publishing. He quotes Reed Elsevier's CEO Crispin Davis, who mentioned in 2003 that ``if you are receiving potential payments for every article submitted, there is an inherent conflict of interest that could threaten the quality of the peer review system''\footnote{Oral evidence to UK House of Commons Science \& Technology Inquiry, March 1st 2004, Sir Crispin Davis (CEO, Reed Elsevier), see \url{https://publications.parliament.uk/pa/cm200304/cmselect/cmsctech/uc399-i/uc39902.htm}}.

The somewhat controversial term ``predatory'' was coined by Jeffrey Beall in his 2010 review \cite{beall2010a} of nine Gold OA scholarly publishers. As a result of his subsequent efforts on maintaining his famous lists, it became a standard term used by notable peer-reviewed articles published in mainstream journals in recent years \cite{shen2015a, xia2015a, shamseer2017a}. Beall himself notes that ``these publishers are predatory because their mission is not to promote, preserve, and make available scholarship; instead, their mission is to exploit the author-pays, Open-Access model for their own profit'' \cite{beall2010a}.

Jeffrey Beall faced a backlash and legal threats after making his lists public, and the term became ``potential, possible or probable predatory''. Both \citet{shen2015a} and \citet{xia2015a} use parentheses largely abstaining from the discussion of the criteria developed by Beall. They do not clarify whether the journals they have studied are really predatory. We follow this practice and use the term ``potentially predatory'' indicating that we use Beall's lists with caution and that the quality of editorial processes in the journals differs widely. 
It should be noted that Beall's approach, which pairs Gold OA as a whole with ``broken'', anything-goes peer review, was frequently criticized mostly by members of the academic librarian community \cite{berger2015a, crawford2014a, esposito2013a}. This discussion came to an end with the unexpected disappearance of Beall's List itself. In January 2017, the author deleted it without explanation. However, in June 2017, Beall summarized his experience about predatory publishing and explained the List disappearance \cite{beall2017a}.

The key issue with PPJ, \ie perceived lack of adequate peer review, is the most difficult to study as the absolute majority of Beall's-listed PPJ do not use an open peer review model. Usually it is studied experimentally, with a fake paper composed manually \cite{bohannon2013a} or using special software like SciGen \cite{davis2009a}.

A recent peer-reviewed article aiming to characterize PPJ \cite{shamseer2017a} compared Beall's-listed PPJ with presumably legitimate open-access and subscription journals. After surveying several hundreds of such venues, the authors came up with a set of 13 features that distinguish PPJ. These, for example, include having a name similar to a known legitimate journal, presenting pseudo-metrics like Global Impact Factor on their websites, and offering rapid peer review and publication. They found that nearly 75\% of PPJ had editors or editorial board members whose affiliation with the journal was unverified while the same measure was 2\% for open access journals and 1\% for subscription-based journals. As for the geography of PPJ, \citet{shamseer2017a} have found that India dominates as a country hosting 40\% of the surveyed PPJ. By contrast, the UK dominates the open access group with 34\% of journals and the USA dominates the subscription-based group with 66\% of journals.

A well-known and widely cited study of PPJ \cite{shen2015a} takes a longitudinal approach to studying a subset of 613 Beall journals sampled from a much larger set constructed by the authors. One of the main results relevant for our study shows that India is a market leader hosting 27.1\% of publishers that produce more than one journal. After studying a sample of 262 corresponding authors, Shen and Björk found that more than 60\% of them are from Asia (specifically, 34.7\% of the studied corresponding authors are from India) and 16.4\% from Africa (8\% from Nigeria). The paper concludes that the ``problem of predatory open access seems highly contained to just a few countries''. Disciplinary structure is also considered, the top three subject areas are ``General'', ``Engineering'' and ``Medicine'' respectively.

Young and inexperienced researchers from developing countries are the core  authors of papers in ``predatory'' journals \cite{xia2015a}. Similarly to \citet{shamseer2017a}, the authors of this paper considered three groups of journals in Biomedical science. The first group of journals consisted of seven journals from Beall's list. The second group of journals included five journals that rejected Bohannon's fake paper \cite{bohannon2013a}. The third group contained high-status OA journals from the PLoS (Public Library of Science)\footnote{See \url{https://www.plos.org/}} series. Affiliations, publication counts, and citation data for authors in three groups of journals were analyzed. As a result, it was found that researchers with less than 5 publications from developing countries like Algeria, Bangladesh, Brazil, Egypt, India, Indonesia, Iran, Nigeria and others prevail among those publishing in the surveyed journals from Beall's list.

Overall, we agree with \citet{xia2017a} that current research on the scale and dynamics of PPJ publishing in mainstream peer-reviewed journals is limited. The vast majority of articles on this topic are opinion pieces about the dangers of predatory publishing for modern academia, and/or checklists or advices for prospective authors, usually in a specific subject area \cite{eriksson2016a, eriksson2017a, balehegn2017a}.

\hl{A contemporary study of the titles delisted by Scopus \cite{cortegiani2020inflated} has shown that journals discontinued for ``publication concerns'' continue to be cited despite discontinuation and predatory behaviour seemed common. These citations may influence scholars' metrics prompting artificial career advancements, bonus systems and promotion.}

Another work \cite{ibba2017a} analyzed the ratio of the number of publications in PPJ and the number of publications in reputable journals in computer science from 2011 to 2015. They used 15 Google Scholar queries with selected key words. They concluded that the portion of publications in PPJ increased between 2011 and 2014, and significantly decreased in 2015 with respect to 2014. Only 6 of 89 identified publishers are indexed in Scopus: 35 journals from Academic Journals and 17 journals from WSEAS publishers.

There are also the results of our previous research project \cite{sterligov2016a}. We matched Beall's lists of journals and publishers to journal titles and publisher names from the Scopus' Source title list. Inclusion of Frontiers Media S.A. publisher in the original Beall's list has been controversial. The prevailing academic view is that the publisher has a sufficient quality \cite{bloudoff-indelicato2015a}. So we excluded 29 sources of Frontiers Media publisher from the list.
Moreover, journals discontinued by Scopus were added.
Hence, the final list included 531 PPJ. 
According to country-level data the most affected countries were Kazakhstan with about 30\% publications in PPJ in 2015 and 47-49\% in 2013-2014 and Indonesia with 23\% in 2015. Few months later we \cite{savina2016a} modified our matching algorithm and identified 665 journals total (447 ``active'' and 218 ``inactive'' journals) that included 29 journals published by Frontiers Media S.A. As a result, we compared publication count during 2011-2015 period for the entire list and the list without Frontiers Media journals. The most affected countries by largest share of PPJ publications in 2015 were Kazakhstan, Indonesia, Iraq, India, Morocco, Malaysia, Nigeria.

Later, \citet{machacek2017a} also used Beall's list to identify 405 PPJ in Scopus. They used Ulrichsweb\footnote{See \url{http://ulrichsweb.serialssolutions.com/}} to extract ISSN for journals in Beall's lists of journals and publishers. This resulted in the omittance of PPJ not mentioned in Ulrichsweb. Machacek and Srholec do not provide the list of PPJ that they have identified in Scopus. They recently updated their results considering three groups of PPJ: standalone journals from Beall's list, 29 journals of Frontiers Research Foundation and journals extracted from Beall's list of publishers \cite{machacek2019a}. Total number of individual journals was 324. According to their results the most affected countries were in Asia and Africa. In particular, Kazakhstan, Indonesia, Iraq and Albania are among them, which is consistent with our prior work \cite{sterligov2016a, savina2016a}.

This study continues our previous work \cite{sterligov2016a}. In this paper, we consider a wider range of issues besides the country distribution, from the scientometric characteristics of the journals and subject areas to the governmental policies. We use three non-intersecting sets of PPJ based on both Beall's lists of journals and publishers as well as sources discontinued by Elsevier. It allows us to draw conclusions and form hypotheses regarding both the scientific level of journals (through citation indicators) and the extent of the PPJ problem. We focus on the research areas that seem extremely relevant in connection with the noted growing role of Scopus in the management of science worldwide. In addition, we specifically consider the case of Kazakhstan and Russia to form a hypothesis about the causal link between the growth of PPJ publications and state policy in research evaluation. We provide the list of potentially predatory journals with their ISSNs in supplementary materials, see table~\ref{tab:list_ppj}.

\section{Methods}\label{sec:methods}

\hl{For the purpose of this study we considered a journal as \textit{potentially predatory} if it is either in Beall's list of standalone journals \cite{beall2016journals}, or its publisher is from Beall's list \cite{beall2016publishers}, or discontinued by Scopus for ``Publication Concerns''}\footnote{Elsevier's site \url{https://www.elsevier.com/solutions/scopus/how-scopus-works/content}}.

\subsection{Data Sources}
We downloaded \textit{Beall's list} of ``potential, possible, or probable predatory scholarly open-access'' publishers and standalone journals \cite{beall2016publishers, beall2016journals}. It consisted of 1,064 publishers and 1,132 journals.
We downloaded Scopus \textit{Discontinued Source list}\footnote{Accessed in September 2018}. This list included sources that were discontinued for three reasons: Metrics, Publication Concerns, and Radar\footnote{See \url{https://www.elsevier.com/solutions/scopus/how-scopus-works/content/content-policy-and-selection}}. We used only 289 sources with tag ``Publication Concerns''.
We downloaded two versions of Scopus \textit{Source title list}\footnote{Elsevier's site \url{https://www.elsevier.com/solutions/scopus/how-scopus-works/content}} in August 2016 and in September 2018.
There are different types of sources included in Source title list: Journal, Trade Journal, Book Series, and Conference Proceedings. We analysed only type ``journal''.

\subsection{Data Generating Process}\label{sec:data generation}

We formed a list of Potentially Predatory Journals (PPJ) in the following way. We first formed a list of potentially predatory sources and then we selected journals from it. 

\begin{figure*}[!ht]
\includegraphics[width=\linewidth]{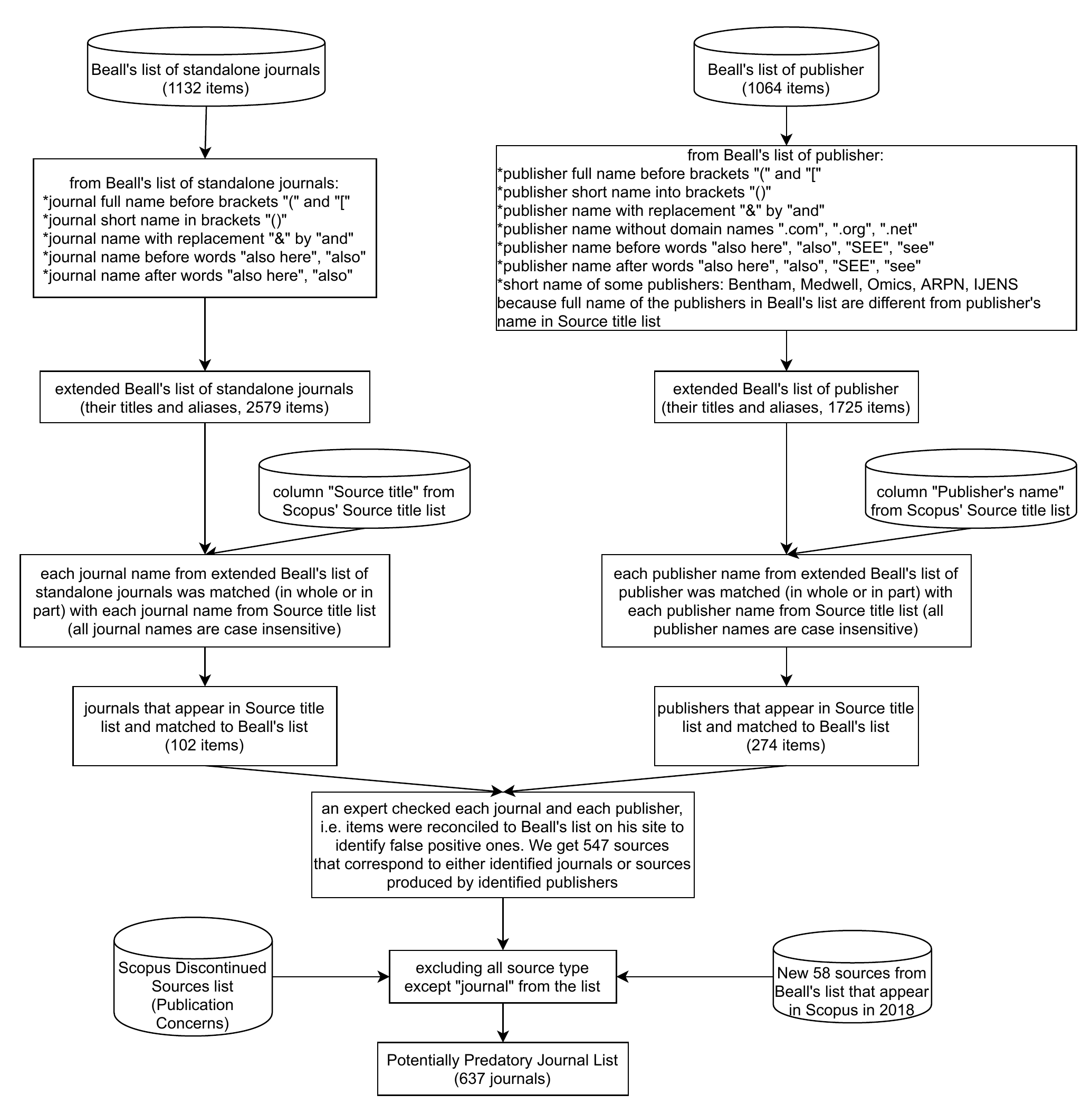}
\caption{Algorithm of formation of list with potentially predatory journals. The algorithm matches journal titles and publisher names from Beall's lists to Source title list in Scopus.\label{fig:diagram_ppj}}
\end{figure*}

We matched journal and publisher titles from Beall's list to source title and publisher's name in the official Scopus Source title list\footnote{August 2016 version}, see Figure \ref{fig:diagram_ppj}. Two titles are considered matched if either title is a substring of another.

We checked the matched items manually to remove false positives. 
For example, Beall's list included Bentham Open publisher and did not include Bentham Science Publishers. The Source title list attributed Bentham Open journals to Bentham Science because Bentham Open was a division of Bentham Science \cite{beall2009a}. So we manually selected only 117 journals published by Bentham Open for our list of potentially predatory sources.
Also, we excluded 29 journals of Frontiers Media publisher from our list.

In September 2018, we identified 58 sources from Beall's list that were included in the Source title list after August 2016. We also identified 78 new sources with the tag ``Publication Concerns'' in the Discontinued Source list (the rest 211 sources were already included in our list). 

For the subsequent analysis, only sources of type ``journal'' were considered. 
The final list of PPJ included 637 titles. The list was divided into the following three non-intersecting groups:

\vspace{1pt}\noindent\textbf{Publication Concerns:} \hl{potentially predatory} journals with tag ``Publication Concerns'' in the Discontinued Source list in September 2018 (252 titles, of which four journals were not in the Source title list).

\vspace{1pt}\noindent\textbf{Active PPJ:} \hl{potentially predatory} journals marked Active (\ie not discontinued) in the Source title list in September 2018 version (215 items). 

\vspace{1pt}\noindent\textbf{Inactive PPJ:} all other \hl{potentially predatory} journals that were marked Inactive in the Source title list in September 2018 version. We also added journals that were in the list in August 2016 but not in September 2018. The total number of journals were 170.

\subsection{Data Analysis}

\hl{In this study we consider only journal articles and reviews as publications.}
We downloaded publication counts in the potentially predatory journals (PPJ) broken down by country, subject area, year and the group of PPJ (see Section \ref{sec:data generation}) from Scopus for the period from 2001 to November 2018. The country of a publication is determined by the affiliation of the authors. So, one publication can be attributed to more than one country. Similarly, a publication can be attributed to multiple subject areas. We used Scopus active journals (excluding PPJ) as a reference group, and downloaded the same publication counts for them. \hl{The reference group had 23362 journals.}

For the analysis of journal metrics, the values for 2016 were selected and taken from the Source title list (September 2018 version) because more journals had these indicators computed for this year compared to other years.

\section{Results}
\subsection{Citation Metrics}

The most important bibliometric characteristic of scholarly journals is their citation impact indicators. There is a plethora of the indicators \cite{waltman2016a}, but none are universal for all use cases. We study diverse journals that represent many different fields of research, thus the normalized metrics are more appropriate. Scopus offers two of them: Source-Normalized Impact per Paper (SNIP) \cite{waltman2013a} and Scimago Journal Rank (SJR) \cite{guerrero-bote2012a}.

\begin{figure*}[!h]
\begin{tabular}{ccc}
   \includegraphics[width=0.45\linewidth]{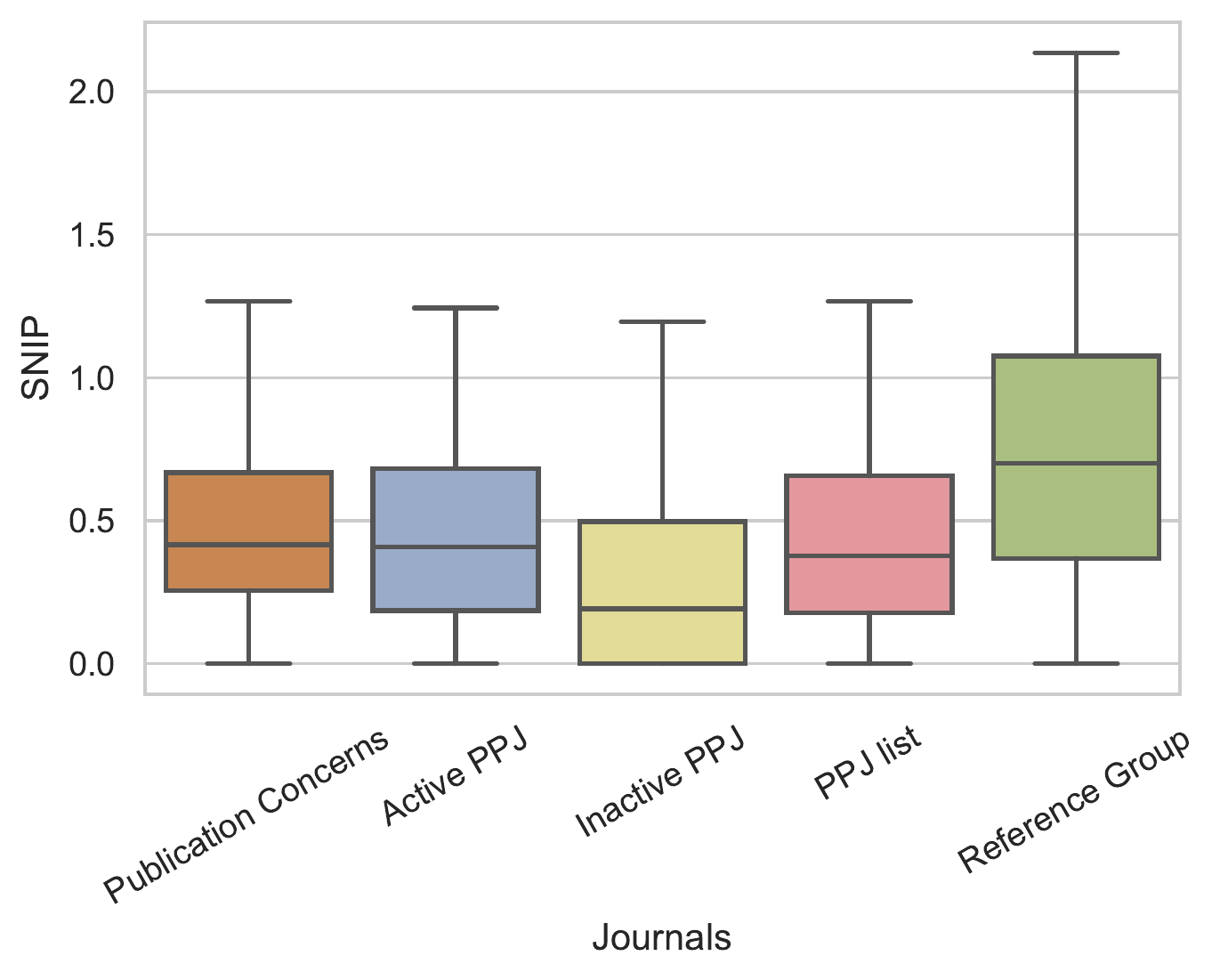} && \includegraphics[width=0.45\linewidth]{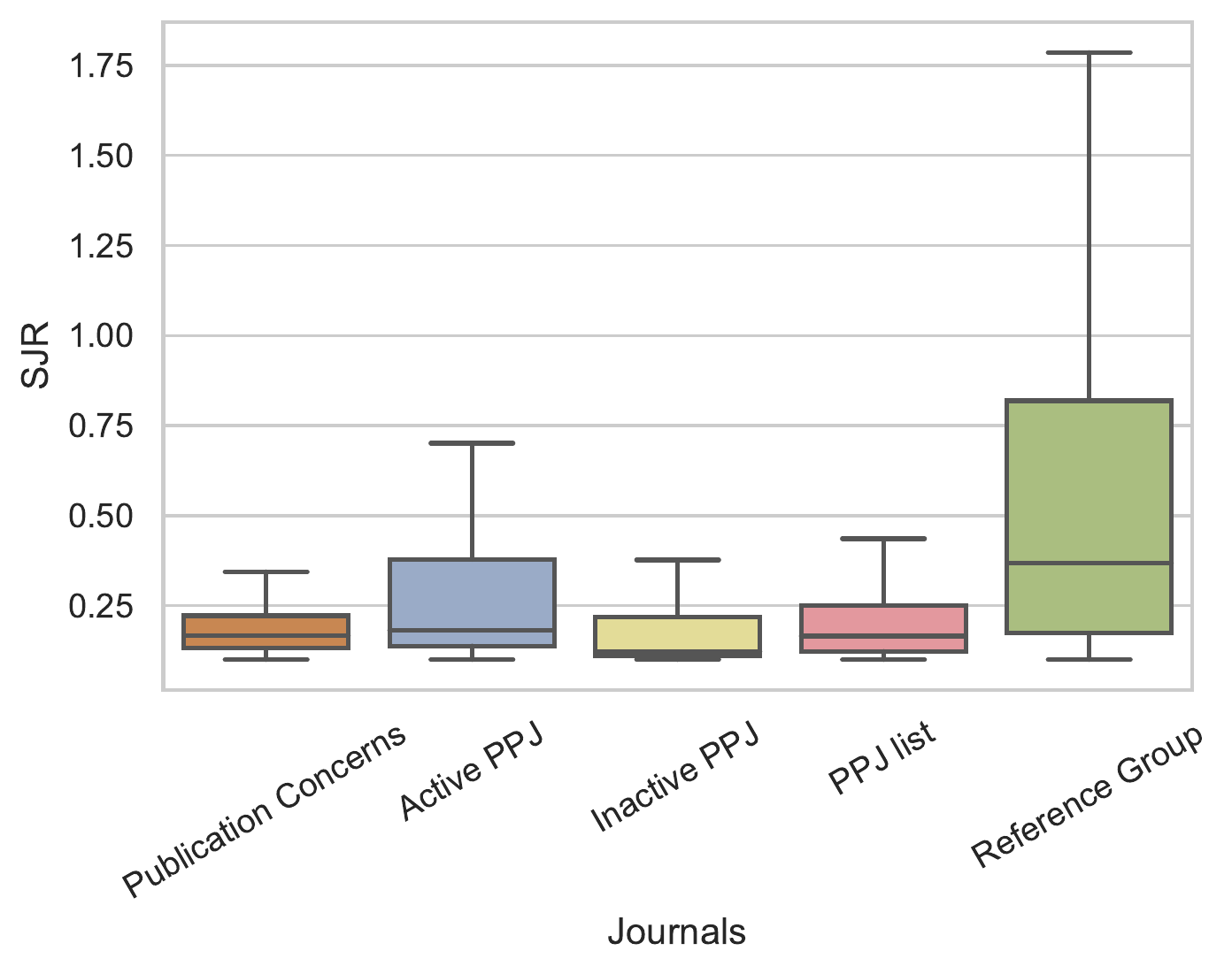} \\
   \parbox{0.45\linewidth}{\caption{\label{fig:snip_2016_boxplot} SNIP value for Publication Concerns group of journals, Active PPJ group, Inactive PPJ group, PPJ list, and Reference Group in 2016}} &&
   \parbox{0.45\linewidth}{\caption{\label{fig:sjr_2016_boxplot} SJR value for Publication Concerns group of journals, Active PPJ group, Inactive PPJ group, PPJ list, and Reference Group in 2016}}
\end{tabular}
\end{figure*}

To identify characteristics of PPJ, the descriptive statistics of journal measures (SNIP, SJR) were considered for the \hl{four} groups of journals: our PPJ list, Publication Concerns, Active PPJ, active journals from the Source title list without potentially predatory journals (reference group), see Table \ref{tab:descr_stat} \hl{and Figures~\ref{fig:snip_2016_boxplot}, \ref{fig:sjr_2016_boxplot}}. In each group of journals, except Publication Concerns journals and all PPJ, we consider only ``active'' journals as of September 2018. For some journals these measures are not available and that means there were missing observations.

\begin{table*}[!h]
\caption{Descriptive statistics for journal metrics}
\label{tab:descr_stat}
\centering
\begin{tabular}{|c|c|r|r|r|r|r|r|r|r|}
\hline
\multirow{2}[4]{*}{2016} & \multirow{2}[4]{*}{} & \multicolumn{2}{p{4.455em}|}{Publication Concerns Journals} & \multicolumn{2}{p{4.455em}|}{Active PPJ} & \multicolumn{2}{p{4.73em}|}{PPJ list (Pub Con + Active PPJ + Inactive PPJ)} & \multicolumn{2}{p{6.18em}|}{Source title list minus PPJ list (``active'' journals)} \bigstrut\\
\cline{3-10}  &   & \multicolumn{1}{p{2.455em}|}{SNIP} & \multicolumn{1}{p{2em}|}{SJR} & \multicolumn{1}{p{2.455em}|}{SNIP} & \multicolumn{1}{p{2em}|}{SJR} & \multicolumn{1}{p{2.365em}|}{SNIP} & \multicolumn{1}{p{2.365em}|}{SJR} & \multicolumn{1}{p{3.09em}|}{SNIP} & \multicolumn{1}{p{3.09em}|}{SJR} \bigstrut\\
\hline
\multicolumn{2}{|p{8.09em}|}{Mean} & 0.54 & 0.2 & 0.49 & 0.3 & 0.49 & 0.25 & 0.83 & 0.72 \bigstrut\\
\hline
\multicolumn{2}{|p{8.09em}|}{Median} & 0.42 & 0.2 & 0.41 & 0.2 & 0.38 & 0.17 & 0.7 & 0.37 \bigstrut\\
\hline
\multicolumn{2}{|p{8.09em}|}{Std. Deviation} & 0.53 & 0.2 & 0.4 & 0.4 & 0.49 & 0.26 & 0.93 & 1.32\bigstrut\\
\hline
\multicolumn{2}{|p{8.09em}|}{Maximum} & 4.56 & 1.3 & 2.44 & 2.5 & 4.56 & 2.46 & 68.18 & 43 \bigstrut\\
\hline
\multicolumn{1}{|c|}{\multirow{2}[4]{*}{Percentiles}} & 25 & 0.26 & 0.1 & 0.19 & 0.1 & 0.18 & 0.12 & 0.37 & 0.17 \bigstrut\\
\cline{2-10}  & 75 & 0.67 & 0.2 & 0.68 & 0.4 & 0.66 & 0.25 & 1.08 & 0.82 \bigstrut\\
\hline
\end{tabular}%
\end{table*}

\begin{figure}[!h]
\includegraphics[width=\linewidth, clip, trim=2cm 16cm 2cm 2cm]{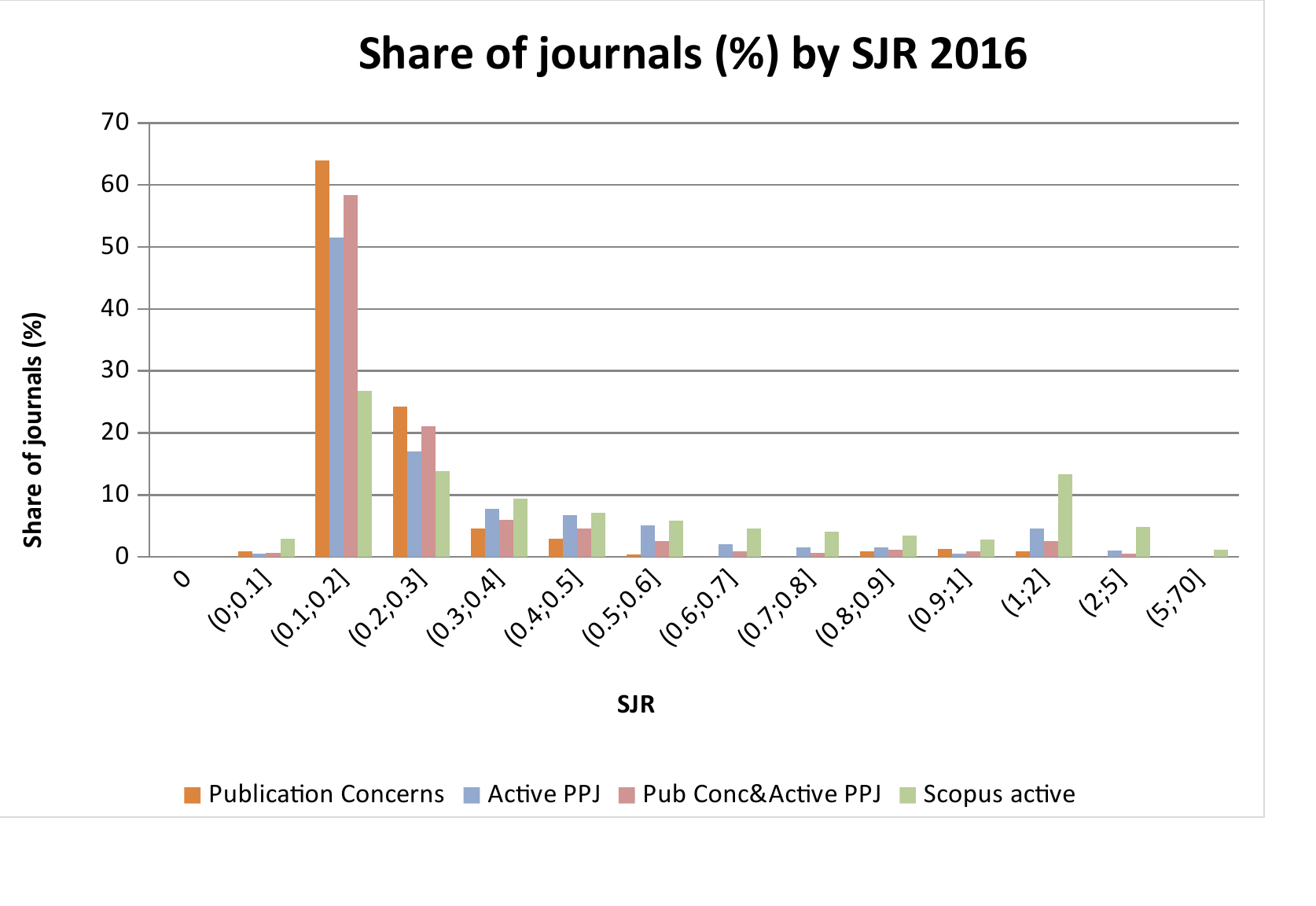}
\caption{\label{fig:sjr_2016} The shares of journals (\%) for each interval of SJR value in 2016. For example, the interval $(0.2;0.3]$ includes journals with the SJR value greater $0.2$ and less than or equal $0.3$.
The figure shows the comparison of the SJR value for the Publication Concerns group, Active PPJ group and  all Scopus active journals (including PPJ). Most of PPJ have the SJR value less than 1. In particular, 58\% of Publication Concerns and Active PPJ have SJR value between 0.1 to 0.2. In contrast, only 27\% of Scopus active journals have the same value.}
\end{figure}

The average and median values of SNIP and SJR for ``active'' journals in PPJ list are lower than for other groups.
About 52\% of Active PPJ and 58\% of Publication Concerns \& Active PPJ with SJR value (without missing values) have SJR value between 0.1 to 0.2 which is compared to about 27\% of all active journals with SJR value in the Source title list with similar values (see Figure \ref{fig:sjr_2016}). There are two active PPJ for which SJR value is more than 2. For instance, ``Aging'' (Impact Journals) has SJR 2016 of 2.458, and ``Open Bioinformatics Journal'' (Bentham Open) has SJR 2016 of 2.384. There are no PPJ titles with SJR value more than 5.

\begin{figure}[!h]
\includegraphics[width=\linewidth, clip, trim=2cm 16cm 2cm 2cm]{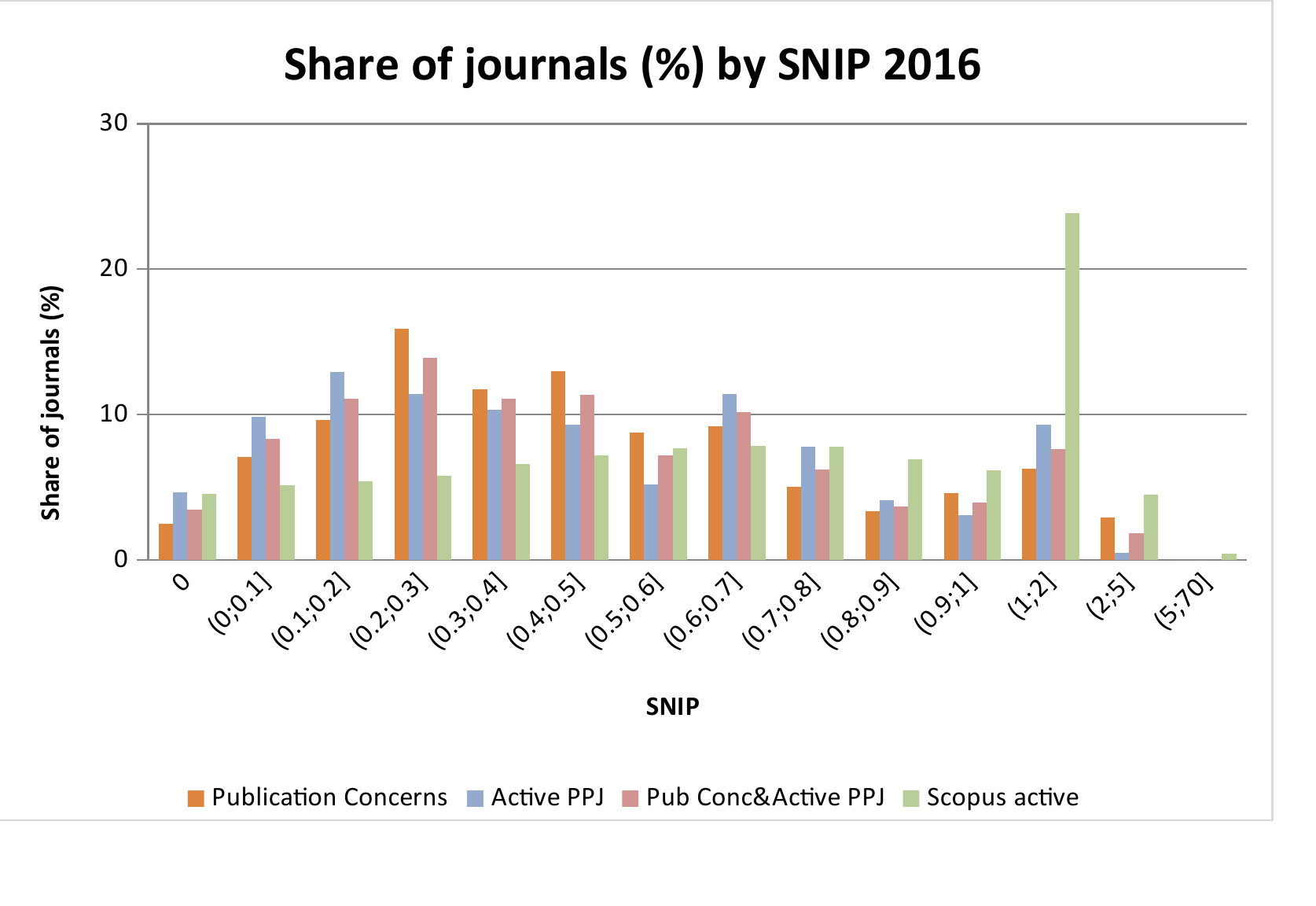}
\caption{\label{fig:snip_2016} The shares of journals (\%) for each interval of SNIP value in 2016.}
\end{figure}

The distribution of SNIP is flatter than the distribution of SJR for Active PPJ group, see Figure \ref{fig:snip_2016}.

We tested the hypothesis if there were statistically significant differences between journal metrics for three groups of journals (Publication Concerns, Active PPJ, Inactive PPJ). 
There were statistically significant differences in the journal metrics depending on the journal group with a significance level of 5\%, see Appendix \ref{app:kruskel}.

The distribution of these indicators can be presented using the box plot, see Figures \ref{fig:snip_2016_boxplot} and \ref{fig:sjr_2016_boxplot}. The average value of SNIP for the Publication Concerns and Active PPJ groups were greater than the average values of SNIP for the Inactive PPJ group. The average value of SJR for the Publication Concerns, Active PPJ and Inactive PPJ groups were similar.

\subsection{Country-Level Data}\label{sec:country}

\begin{figure}[!h]
\includegraphics[width=\linewidth, clip, trim=1.8cm 17.3cm 2cm 2cm]{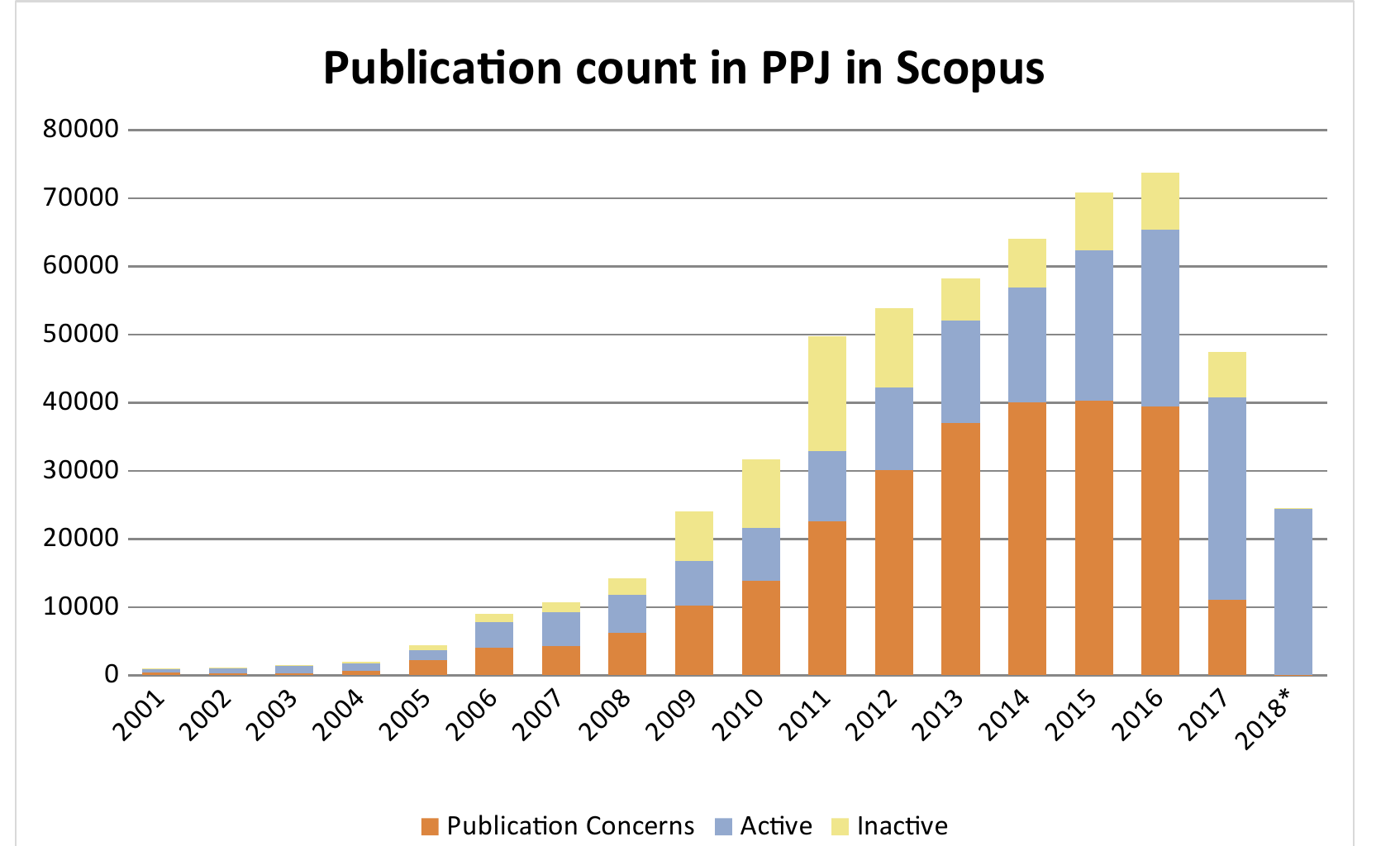}
\caption{\label{fig:ppj_world} Publication counts in PPJ in Scopus. 
Scopus used to stop indexing journals in inactive and publication concern groups at the time of inclusion of such journals to these groups. As our Inactive and Publication Concerns lists were based on the 2018 data, there were virtually no publications in these groups in 2018. Specifically, only a few publications in these groups were still indexed, \ie 44 and 159 in Publication Concerns and Inactive groups respectively.
*The plot was based on data downloaded in November 2018.}
\end{figure}

The worldwide distribution of the publication count for three groups of PPJ in Scopus is presented in Figure \ref{fig:ppj_world}. The number of publications in PPJ has been increasing significantly since 2011. The peak of publications number in PPJ in Scopus was in 2016. The number of publications in the Publication Concerns journal group decreased in 2017 because Scopus delisted 140 journals.

To analyse publications in PPJ, we selected 76 countries for which the number of publications (articles and reviews) in all journals was more than 10,000 for the period 2011-2018\footnote{The publication counts data were accessed in November 2018. So, the analysis for 2018 was based on preliminary data.}. The United States and China had the largest total number of publications in Scopus during the same period, see Table~\ref{tab:pub_count}.

\begin{table*}[!h]
\caption{Publications count (articles and reviews in Scopus-indexed journals) and PPJ share for countries for the period 2011-2018. *The table was based on data accessed in Nov 2018.}
\label{tab:pub_count}
\centering
\begin{tabular}{|l|l|l||l|l|l|}
\hline
Country       & Count   & PPJ share & Country    & Count & PPJ share \\
\hline
United States      & 3,551,810 & 0.68\%    & Pakistan        & 86571 & 6.98\%    \\
China    & 2,834,181 & 3.32\%    & Argentina       & 85,883 & 0.51\%    \\
United Kingdom     & 1,065,688 & 0.47\%    & Thailand        & 78,000 & 6.38\%    \\
Germany       & 945,778  & 0.43\%    & Romania    & 77,554 & 3.51\%    \\
India    & 749,684  & 13.00\%   & Ireland    & 76,208 & 0.47\%    \\
Japan    & 733,283  & 0.98\%    & Chile      & 73,013 & 0.62\%    \\
France        &668,42& 0.50\%    & Hungary    &65,308& 0.58\%    \\
Italy    &596,605& 1.22\%    & Ukraine    &61,706& 5.26\%    \\
Canada        &589,023& 0.52\%    & Colombia        &52,823& 1.87\%    \\
Spain    &534,593& 0.54\%    & Serbia     &47,387& 4.22\%    \\
Australia     &533,209& 0.63\%    & Indonesia       &45,113& 26.11\%   \\
South Korea   &491,152& 4.69\%    & Nigeria    &42,289& 17.18\%   \\
Brazil        &432,373& 1.14\%    & Croatia    &41,798& 2.45\%    \\
Russia &379,282& 4.43\%    & Slovakia        &39,631& 2.83\%    \\
Netherlands   &340,961& 0.33\%    & Slovenia        &38,368& 0.94\%    \\
Iran     &323,73& 9.15\%    & Tunisia    &37,437& 2.96\%    \\
Turkey        &259,591& 4.12\%    & Algeria    &30,863& 8.05\%    \\
Switzerland   &257,543& 0.34\%    & Vietnam        &27,42& 2.77\%    \\
Poland        &246,072& 0.65\%    & Morocco    &26,78& 13.31\%   \\
Sweden        &228,845& 0.51\%    & Bulgaria        &24,589& 2.42\%    \\
Taiwan        &228,489& 2.68\%    & UAE &22,522& 4.35\%    \\
Belgium       &191,202& 0.45\%    & Lithuania       &20,917& 1.21\%    \\
Denmark       &151,456& 0.38\%    & Bangladesh      &20,83& 8.90\%    \\
Malaysia      &146,452& 17.51\%   & Jordan     &18,557& 13.16\%   \\
Austria       &133,763& 0.52\%    & Iraq       &16,847& 22.27\%   \\
Mexico        &128,33& 1.49\%    & Estonia    &16,782& 0.53\%    \\
Portugal      &126,155& 0.72\%    & Kenya      &16,544& 3.02\%    \\
Czech Republic     &122,441& 1.64\%    & Cuba       &14,572& 0.76\%    \\
South Africa  &121,238& 4.09\%    & Qatar      &14,402& 1.84\%    \\
Israel        &119,169& 0.54\%    & Philippines     &14,119& 5.89\%    \\
Norway        &115,907& 0.49\%    & Ethiopia        &13,858& 6.53\%    \\
Saudi Arabia  &112,286& 7.89\%    & Lebanon    &13,465& 2.44\%    \\
Finland       &110,643& 0.41\%    & Kazakhstan      &13,319& 30.65\%   \\
Singapore     &107,909& 0.82\%    & Peru       &12,072& 0.86\%    \\
Egypt    &106,544& 11.07\%   & Belarus    &11,722& 0.90\%    \\
Greece        &103,3& 1.64\%    & Venezuela       &11,622& 1.47\%    \\
Hong Kong     &100,248& 0.89\%    & Cyprus     &11,592& 2.10\%    \\
New Zealand   &88,861& 0.48\%    & Ghana      &11,52& 5.48\%    \\
\hline
\end{tabular}
\end{table*}

Countries with the largest number of publications in PPJ during the period 2011-2018 were India with 97,454 publications and China with 94,058. The next top-8 countries with the largest numbers of publications in PPJ in the period 2011-2018 were Iran with 29,616 publications, Malaysia with 25,637, the United States with 24,064, South Korea with 23,017, Russia with 16,812, Egypt with 11,796, Indonesia with 11,781, and Turkey with 10,688 publications.
Table \ref{tab:country} shows the change of publication shares (\%) in PPJ over time.

\begin{table*}[!h]
\caption{Dynamics of percentage shares of PPJ publications for top-15 countries from 2010 to 2018 sorted by values in 2016. *The table was based on data downloaded in November 2018.}
\label{tab:country}
\centering
\begin{tabular}{|l|l|l|l|l|l|l|l|l|l|}
\hline
Share PPJ (\%) & 2010  & 2011  & 2012  & 2013  & 2014  & 2015  & 2016  & 2017  & 2018*\bigstrut\\
\hline
Kazakhstan     & 2.22  & 6.35  & 17.43 & 49.02 & 47.39 & 35.97 & 41.61 & 18.66 & 7.88  \bigstrut\\
\hline
Indonesia      & 11.12 & 16.25 & 12.82 & 17.65 & 28.32 & 32.87 & 34.99 & 30.40 & 18.16 \bigstrut\\
\hline
Iraq           & 23.38 & 37.06 & 24.51 & 20.53 & 19.85 & 19.05 & 28.52 & 14.68 & 23.49 \bigstrut\\
\hline
Malaysia       & 20.15 & 26.59 & 20.29 & 20.12 & 17.10 & 17.30 & 19.77 & 13.59 & 9.09  \bigstrut\\
\hline
Morocco        & 6.45  & 8.85  & 13.06 & 14.23 & 17.63 & 17.08 & 18.45 & 12.16 & 3.66  \bigstrut\\
\hline
India          & 10.68 & 14.77 & 13.92 & 13.67 & 16.10 & 18.12 & 15.18 & 6.62  & 6.32  \bigstrut\\
\hline
Egypt          & 8.51  & 14.24 & 17.51 & 14.98 & 11.81 & 11.39 & 12.80 & 5.51  & 4.25  \bigstrut\\
\hline
Jordan         & 17.21 & 19.74 & 21.16 & 16.10 & 14.63 & 11.15 & 12.57 & 8.39  & 5.86  \bigstrut\\
\hline
Nigeria        & 35.08 & 34.00 & 28.42 & 25.85 & 19.02 & 15.18 & 11.56 & 5.32  & 4.09  \bigstrut\\
\hline
Algeria        & 10.70 & 11.23 & 9.57  & 10.19 & 10.00 & 11.26 & 10.95 & 3.29  & 1.89  \bigstrut\\
\hline
Iran           & 12.68 & 21.02 & 15.12 & 12.49 & 8.40  & 7.48  & 8.90  & 3.82  & 1.90  \bigstrut\\
\hline
Philippines    & 2.19  & 3.76  & 3.40  & 2.92  & 8.61  & 10.64 & 8.67  & 5.04  & 2.06  \bigstrut\\
\hline
Russia         & 0.24  & 0.37  & 0.69  & 3.66  & 6.01  & 8.23  & 8.41  & 3.19  & 1.95  \bigstrut\\
\hline
Saudi Arabia   & 10.21 & 12.37 & 11.41 & 12.62 & 9.68  & 7.39  & 7.42  & 4.74  & 3.07  \bigstrut\\
\hline
South Korea    & 1.20  & 1.73  & 2.95  & 4.64  & 5.93  & 6.77  & 7.24  & 5.19  & 1.64\bigstrut\\
\hline
\end{tabular}
\end{table*}

The maximum number of publications in PPJ in Scopus was in 2016. Kazakhstan had the largest share (\%) of such publications, \ie  41.61\%. Indonesia had the second largest share of publications in PPJ, \ie 34.99\%. The share of publications in PPJ increased in 2016 compared to 2011 for the following countries: Kazakhstan, Indonesia, Morocco, Philippines, Russian Federation, Ukraine, South Korea, Italy, Slovakia, Colombia, and Bulgaria. During the same period, the share of publications in PPJ decreased for the following countries: Iran, Nigeria, Jordan, Bangladesh, Serbia, Turkey.

In contrast, the share of publications in PPJ was less than 1\% for many countries, for example, for the United States, the United Kingdom, Germany, France, Canada and Spain in the period 2010-2018, see Table~\ref{tab:share_ppj}.

\begin{table*}[!h]
\caption{The share of publications (articles and reviews) in PPJ for top-10 countries with largest amount of publications in Scopus in the period 2011-2018. *The table was based on data downloaded in November 2018.}
\label{tab:share_ppj}
\centering
\begin{tabular}{|l|l|l|l|l|l|l|l|l|l|}
\hline
Share PPJ (\%) & 2010 & 2011 & 2012 & 2013 & 2014 & 2015 & 2016 & 2017 & 2018*\bigstrut\\
\hline
The United States & 0.57 & 0.61 & 0.62 & 0.57 & 0.61 & 0.79 & 0.93 & 0.80 & 0.45\bigstrut\\
\hline
China             & 1.44 & 3.09 & 4.64 & 4.24 & 3.81 & 4.08 & 3.75 & 2.87 & 0.72\bigstrut\\
\hline
The United Kingdom & 0.50 & 0.54 & 0.49 & 0.42 & 0.46 & 0.47 & 0.59 & 0.47 & 0.35\bigstrut\\
\hline
Germany           & 0.34 & 0.35 & 0.30 & 0.32 & 0.33 & 0.48 & 0.60 & 0.58 & 0.44\bigstrut\\
\hline
Japan             & 0.75 & 0.93 & 0.92 & 0.94 & 1.04 & 0.98 & 1.14 & 1.02 & 0.80\bigstrut\\
\hline
France            & 0.46 & 0.52 & 0.43 & 0.39 & 0.47 & 0.56 & 0.66 & 0.55 & 0.42\bigstrut\\
\hline
Italy             & 0.78 & 0.85 & 0.94 & 1.04 & 1.15 & 1.36 & 1.66 & 1.54 & 1.04\bigstrut\\
\hline
Canada            & 0.46 & 0.56 & 0.54 & 0.46 & 0.48 & 0.52 & 0.71 & 0.54 & 0.35\bigstrut\\
\hline
Spain             & 0.49 & 0.55 & 0.49 & 0.47 & 0.46 & 0.61 & 0.69 & 0.62 & 0.36\bigstrut\\
\hline
\end{tabular}
\end{table*}

The share of publications in PPJ for China was 2.87\% in 2017, about 2 times greater than in 2010 (1.44\%). Since 2011, the share of publications in PPJ for China was more than 3\% with the maximum value of 4.64\% in 2012. The share of publications in PPJ from China decreased after 2013, see Table~\ref{tab:share_ppj}.
This is aligned with the implementation by the Government of China of harsh policies to regulate China's publishing market, resulting in the suspension of many ``trash'' journals \cite{lin2014a}. In 2018, the Communist Party of China required the ministry of science and technology to establish a blacklist of academic journals \cite{cyranoski2018a}.

The share of publication in PPJ for Iran dropped from 21.02\% in 2011 to 3.82\% in 2017 and then to 1.90\% in 2018. 

\subsection{Subject-Level Data}

The largest subject area is ``Engineering'' which included 93,275 publications in PPJ in the period 2011-2018. The highest increase for ``Engineering'' was in 2015. The second-largest is ``Medicine'' with 85,188 publications during the same period. The third largest category is ``Biochemistry, Genetics and Molecular Biology'' at 65,498 publications. The fourth largest subject field is ``Computer Science'' at 61,699 publications from 2011 to 2018, but the number of publications decreased in 2017 compared with 2012, from 13,782 to 5,865 publications. Category ``Pharmacology, Toxicology and Pharmaceutics'' had more than 50000 publications (55,887 publications) but it had the third largest amount of publications of 11,500 in 2016. For subject area ``Agricultural and Biological Sciences'', the number of publications dramatically declined in 2018 compared with 2011, \ie from 11,014 to 1,332 publications. The number of publications in PPJ increased in 2016 compared to 2011 for four subject categories: ``Mathematics'', ``Chemistry'', ``Social Sciences'', ``Materials Science'' with a total number of publications in the period 2011-2018 between 28,000 to 50,000.

Note that almost one third of publications in PPJ in 2011 was in the subject category ``Multidisciplinary'', but in 2018 the share of publications in that field dramatically decreased to less than 1\%, probably because of delisting of several prominent PPJ titles (for example, ``Indian Journal Of Science And Technology'', ``World Applied Sciences Journal'', ``Journal of Applied Science'').

Table \ref{tab:subject} represents the dynamics of publication share (\%) by top-15 subject area sorted by the number of publications in the period from 2011 to 2018.

\begin{table*}[]
\caption{Percentage share (\%) of PPJ publications by top-15 subject area}
\label{tab:subject}
\centering
\begin{tabular}{|p{0.2\textwidth}|p{0.05\textwidth}|p{0.05\textwidth}|p{0.05\textwidth}|p{0.05\textwidth}|p{0.05\textwidth}|p{0.05\textwidth}|p{0.05\textwidth}|p{0.05\textwidth}|p{0.12\textwidth}|}
\hline
Share of publications in PPJ by subject area (\%) & 2011  & 2012  & 2013  & 2014  & 2015  & 2016  & 2017 & 2018 & Publications in PPJ in 2011 - 2018\bigstrut\\
\hline
Engineering                                  & 3.81  & 3.70  & 3.43  & 4.12  & 6.27  & 4.84  & 3.05 & 2.90  & 93,275\bigstrut\\
\hline
Medicine                                     & 1.21  & 1.21  & 1.14  & 1.57  & 2.15  & 2.84  & 3.24 & 1.59  & 85,188\bigstrut\\
\hline
Biochemistry, Genetics and Molecular Biology & 3.26  & 2.27  & 2.98  & 3.53  & 3.48  & 3.72  & 1.81 & 2.27  & 65,498\bigstrut\\
\hline
Computer Science                             & 7.84  & 13.64 & 9.28  & 6.22  & 4.87  & 5.73  & 4.53 & 5.19  & 61,699\bigstrut\\
\hline
Pharmacology, Toxicology and Pharmaceutics   & 9.63  & 7.90  & 8.90  & 11.38 & 10.50 & 13.28 & 4.82 & 3.99  & 55,887\bigstrut\\
\hline
Mathematics                                  & 5.28  & 7.72  & 5.22  & 4.72  & 5.57  & 5.66  & 4.18 & 1.03  & 42,839\bigstrut\\
\hline
Chemistry                                    & 1.91  & 2.17  & 3.00  & 2.81  & 2.97  & 2.94  & 1.61 & 0.86  & 38,111\bigstrut\\
\hline
Social Sciences                              & 2.09  & 3.54  & 2.68  & 3.53  & 3.89  & 3.24  & 2.32 & 0.36  & 37,745\bigstrut\\
\hline
Agricultural and Biological Sciences         & 6.67  & 3.22  & 2.44  & 2.71  & 1.49  & 0.95  & 0.66 & 0.75  & 34,358\bigstrut\\
\hline
Environmental Science                        & 3.41  & 4.54  & 3.62  & 3.32  & 2.57  & 3.01  & 2.81 & 3.54  & 32,032\bigstrut\\
\hline
Materials Science                            & 2.20  & 1.91  & 1.79  & 1.81  & 2.47  & 2.45  & 1.58 & 0.31  & 28,745\bigstrut\\
\hline
Multidisciplinary                            & 27.29 & 23.51 & 27.13 & 14.88 & 9.92  & 9.31  & 0.25 & 0.21  & 25,779\bigstrut\\
\hline
Economics, Econometrics and Finance          & 5.94  & 5.97  & 8.30  & 12.11 & 12.60 & 9.21  & 6.85 & 1.91  & 24,115\bigstrut\\
\hline
Chemical Engineering                         & 1.57  & 2.01  & 2.56  & 3.32  & 3.46  & 3.21  & 1.77 & 4.36  & 23,561\bigstrut\\
\hline
Physics and Astronomy                        & 1.21  & 0.65  & 0.93  & 0.92  & 1.19  & 1.60  & 0.94 & 0.23  & 17,516\bigstrut\\
\hline
\end{tabular}
\end{table*}

\section{Government Policy Impact}

The differences between countries in the shares of publications in PPJ suggest that the growth of publications in PPJ is associated with government policy. \citet{machacek2019a} formulated such hypothesis, but did not substantiated it. Obviously, the identification of the causal link requires special studies and a detailed analysis of incentives to publish and the configuration of research evaluation systems.
To address the possible existence of such dependence, we consider the case of Kazakhstan as the most affected country according to our results in Section~\ref{sec:country}, and also briefly discuss the case of Russia, suggesting a possible influence of a national university excellence initiative. These examples problematize varied consequences of metric usage.

\subsection{Case of Kazakhstan}

\begin{figure}
\includegraphics[width=\linewidth, clip, trim=1.8cm 6.8cm 1.5cm 8.1cm]{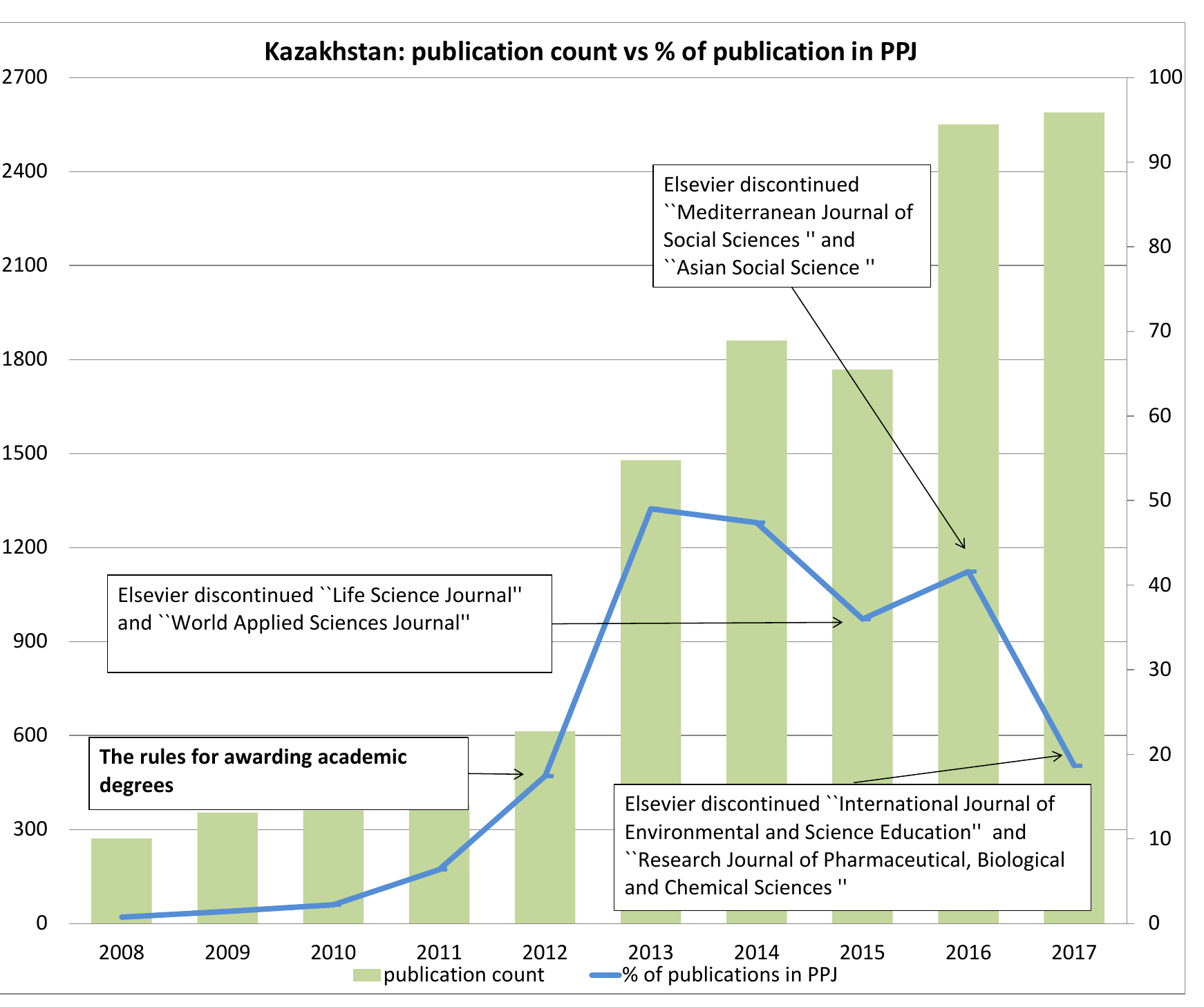}
\caption{\label{fig:ppj_kazakh} Kazakhstan: publication count vs share of publication (\%) in PPJ. The figure demonstrates dynamics of the share of PPJ publications and publication count in Scopus-indexing journals before and after the Rules for awarding academic degrees. The share of publication in PPJ rapidly increased after implementation of the Rules in 2012, that is from 17\% in 2012 to 49\% in 2013.
The bar chart (corresponds to the left axis) shows the growth of publication count in all journals in Scopus from 272 in 2008 to 2,589 publications in 2017. The line (corresponds to the right axis) shows the share of publication in PPJ.}
\end{figure}

In 2011-2012, the government of Kazakhstan implemented new Rules for awarding academic degrees that included a requirement to publish at least one paper in Scopus or WoS-indexed journals, see details in Appendix \ref{app:kaz_new_rules}.
The publications from Kazakhstan in all journals in Scopus increased sharply as well as publications in potentially predatory journals since 2013, see Figure \ref{fig:ppj_kazakh}. We indicate that the rapid growth in PPJ publications for Kazakhstan from 2013 was associated with the new Rules.

In 2014 Elsevier stopped indexing 35 journals from the PPJ list, which were popular among authors from Kazakhstan. For example, more than a quarter of all articles and reviews from Kazakhstan were published in ``Life Science Journal''  and about 5.51\% of all articles and reviews from Kazakhstan were published in “World Applied Sciences Journal” in 2014. Both journals were delisted by Elsevier in 2014 resulting in reduced number of publication in PPJ in 2015.

Elsevier stopped indexing 27 journals from the PPJ list in 2015. 3.9\% of articles and reviews from Kazakhstan were published in the delisted ``Mediterranean Journal of Social Sciences'' and 3.68\% of publications in the discontinued ``Asian Social Science'' in 2015. In 2016, 140 journals from the PPJ list were discontinued. The largest share of Kazakhstan articles and reviews were published in ``International Journal of Environmental and Science Education'' (8.67\%) in 2016.

The new policy encouraged publication activity mostly among young researchers, which is consistent with the results of \citet{xia2015a} showing that most publications in PPJ were attributed to young and inexperienced researchers from developing countries.

The burst of low-quality publications in Kazakhstan is a fact known to local authorities. Their reaction to the problem was further formalization of the list of journals through quantitative indicators and recommending additional ``white lists''. The success of this approach is questionable. For example, there are PPJ among those that satisfy the required threshold\footnote{See the order of the Minister of Education and Science of Kazakhstan from 03/31/2011 No. 127, appendix~1 \url{http://web.archive.org/web/20190905234744/https://egov.kz/cms/ru/law/list/V1100006951}.} of the top 25\% according to the CiteScore.

\subsection{Case of Russia}

\begin{figure}
\includegraphics[width=\linewidth, clip, trim=1.8cm 6.8cm 1.5cm 8.1cm]{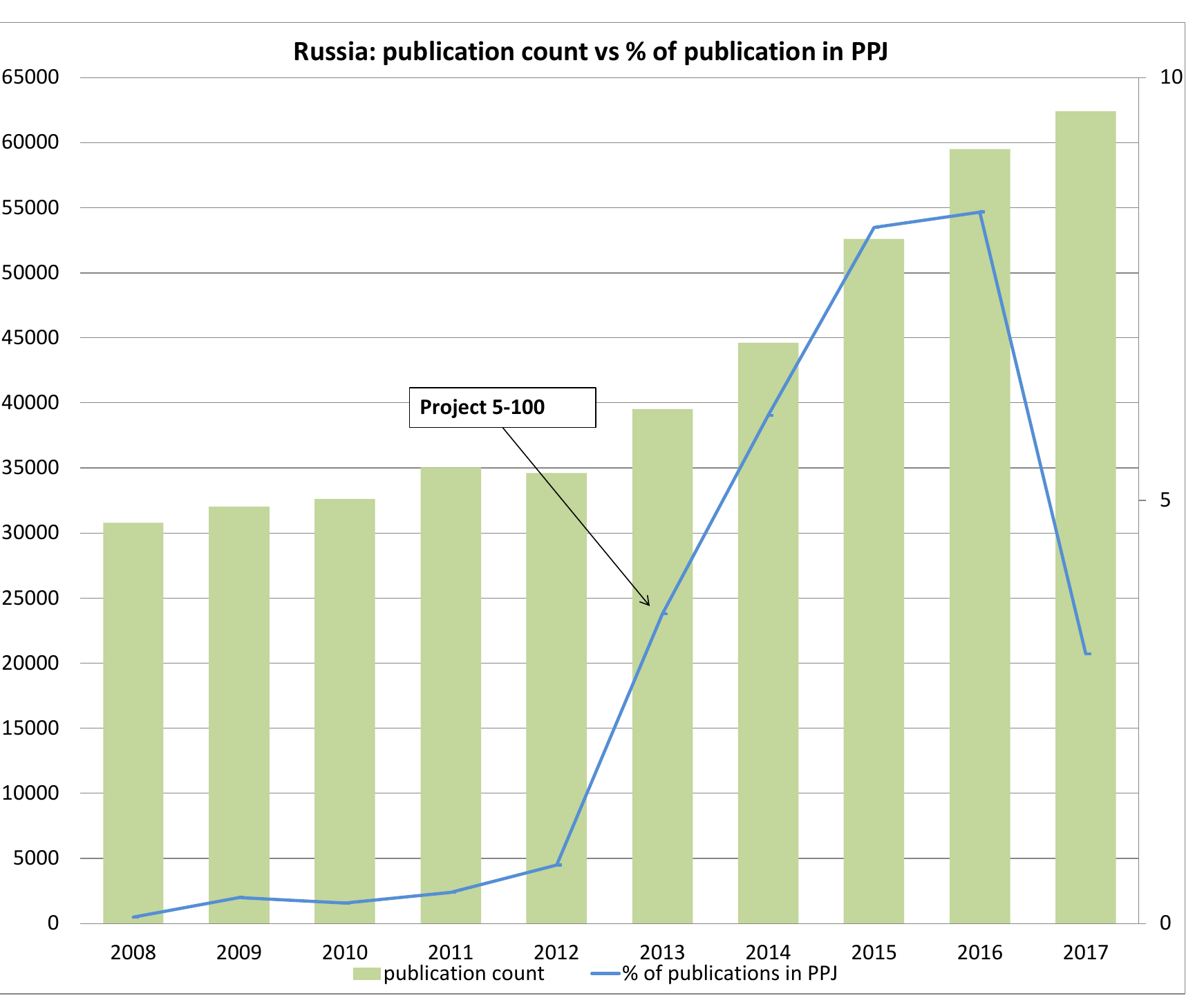}
\caption{\label{fig:ppj_russia} Russia: publication count vs share of publication (\%) in PPJ. The figure demonstrates dynamics of the share of PPJ publications and publication count in Scopus-indexing journals before and after implementing Russian academic excellence project (``Project 5-100''). The bar chart (corresponds to the left axis) shows the growth of publication count in all journals in Scopus from 30,801 in 2008 to 62,406 publications in 2017. The line (corresponds to the right axis) shows the share of publication in PPJ.}
\end{figure}

Russia has seen a huge increase in bibliometric Key Performance Indicators' usage on all levels in recent years. \citet{moed2018a} offer a brief description and discussion of Russian policy, including university reporting requirements, pay-per-publication academic bonuses, PhD and tenure regulations.

One of the most important initiatives was set in accordance with the government's main goals of development, stated in the president's 2012 decrees. Launched in the same year, 5-top-100 project\footnote{\url{https://5top100.ru/en/}} supports select 21 universities, so that at least 5 of them are in the top 100 of international university rankings by 2020. Exact rankings were not mentioned, so it is not clear if the project has been successful, but all participants were to demonstrate a yearly evaluated increase in WoS/Scopus publication counts. Within the course of a few years, there was a surge of PPJ articles from select participating universities, and soon the practice became widespread. Figure~\ref{fig:ppj_russia} shows the resulting significant growth of PPJ paper counts for Russia as a whole.

\hl{\subsection{Other Evidence}

Besides Kazakhstan and Russia, several facts support our hypothesis that predatory publishing is linked to simplistic government policies based on Scopus publication counts. First of all it is the proliferation of various journal blacklists used to determine which publications will not be counted in evaluation formulas \cite{sterligov2020blacklists}. For example, such list, compiled by Malaysian Ministry of Education, is published on local universities' websites\footnote{\url{http://rimc.uum.edu.my/index.php/blacklisted-journals-by-moe}}. Harsh paper-counting policies are observed in Indonesia\footnote{The Jakarta Post article \url{https://www.thejakartapost.com/news/2018/06/10/wanted-6000-new-journals-to-publish-150000-papers.html}: ``The regulation requires academics to publish at least one scientific paper in three years in an international or accredited journal. Another regulation has also contributed to the surge in published papers. Three years ago, the government issued Ministerial Regulation No. 44/2015 on higher education quality, which required every graduate student to publish one piece in an accredited journal and a doctoral candidate to publish a piece in an international journal''}. \citet{mouton2017extent} noted that ``scholarly publishing in South African is strongly influenced by the [Government's] system of paying subsidies to universities for research publications \ldots [this] is the major driver behind the huge increase in publication output since 2005 and has become the major incentive for many academics to publish and publish as many articles as quickly as they can''.
What is important, however, is that at least some countries heavily affected by predatory publishing seem to lack a centralised list-based policy, while paper-counting policies directly linked to career promotions exist on the university level, see \citet{nwagwu2015penetration, ajuwon2018predatory}.}

\section{Discussion and Conclusion}

\subsection{Countries and Subject Areas}

We examined the distribution of publications by subject area according to the All Science Journal Classification (ASJC) in Scopus database. The classification has 27 main fields comprising 334 subjects. Publications in PPJ cover all subjects, most of them are significantly affected. For 19 fields, the number of publications exceeded 10,000 for the 2011-2018 period. The largest number of publications in PPJ during the 2011-2018 period was in “Engineering”. However, the share of such publications is not large compared to the total number of publications in Scopus on this subject. The most affected area in terms of PPJ article share was “Multidisciplinary”, largely due to several predatory “megajournals”, which were subsequently delisted by Elsevier so that PPJ share dropped from 27\% in 2013 to 0.21\% in 2018.

Our analysis of the affected countries is consistent with the results of other studies \cite{xia2015a, shen2015a, sterligov2016a, machacek2019a}, but our dataset includes more journals (hence, 418,040 publications during 2011-2017 period) and uses a longer timeframe. In contrast, we consider a wider range of issues besides the country distribution, \ie the scientometric characteristics of the journals and subject areas, and the governmental policies.

Additionally, we examined the dynamics of publication counts starting from 2001. We showed that a significant increase in the number of publications started in 2007, when the total number of PPJ articles and reviews exceeded 10,000, but this growth pattern widely differs for the surveyed countries. For example, in Russia and Kazakhstan the surge started in 2013, which was primarily connected with the public policy of these countries, while in China and Iran this problem appeared much earlier. The most affected countries were the developing ones: Kazakhstan, Indonesia, India, Nigeria and Egypt, Iraq and Iran, see \cite{machacek2019a} for comparisons of PPJ shares and income of countries.

For Argentina, Australia, Austria, Belgium, Canada, Chile, Denmark, Finland, France, Germany, Hungary, Ireland, Israel, Netherlands, New Zealand, Norway, Poland, Portugal, Spain, Sweden, Switzerland, United Kingdom, and United States the percentage of publications in PPJ is small, less than 1\% for the 2011-2018 period. The only notable exception amongst developed countries is South Korea, where PPJ share increased from 1.73\% in 2011 to 7.24\% in 2016 and then fell to 5.19\% in 2017. Our data suggests that the problems with journal quality are faced usually by the authors from less developed countries, but it is clear that they arise also in the countries with more elaborate and mature academic systems.

Based on the case of Kazakhstan and Russia, we suggest that the growth of PPJ is significantly impacted to government science policy. Although we cannot make claims about a causal connection, the (im)balance of government policy measures, the discontinuation of indexing PPJ popular in Kazakhstan by Elsevier, and the number and the share of publications in PPJ are highly illustrative. Thus it is reasonable to conjecture that the growth of PPJ is affected by the research evaluation policies implementing similar Scopus-based performance indicators, such as publication requirements, university reporting requirements, academic bonuses and tenure policies.

\subsection{State of PPJ in Scopus}

Elsevier self-cleans its Scopus database by publishing a discontinued Source list and updating it. Since most of the delisting occured in 2016, the number of publications by Publication Concerns group fell sharply in 2017 (from nearly 40,000 publications in 2016 to more than 10,000 in 2017). However, the other, active group of surveyed PPJ shows linear growth of the publication counts, which raises many questions. We observe that those who earlier published in PPJ, after their delisting, have quickly switched to publishing in a plethora of new Scopus journals that can be equally dubious, but counted as ``good'' in this and similar studies. We leave the detailed study of this phenomenon for the future. 

It is important to point out Elsevier's interest in addressing this problem. For example, our research team has been interacting with Elsevier's experts on the issue since 2016. 
Previous studies have shown that journal inclusion in a particular data source used for evaluation leads to increased incentives to publish in PPJ \cite{bagues2017a}. \hl{Particularly worrying are the results of a recent citation study \cite{cortegiani2020inflated}, that has revealed that journals discontinued in Scopus for ``publication concerns'' continue to be cited despite discontinuation.} However, Scopus is not the only database affected. Our preliminary studies show that at least several dozens of PPJ already delisted by Elsevier are listed as active in the Web of Science Core Collection (mostly in Emerging Sources Citation Index).

The incentives to publish for the sake of publication or indicators are only increasing \cite{biagioli2020a}, and new ``fake'' publications are no longer so easily detected \cite{biagioli2019a}. 
It is becoming increasingly difficult to deal with the issue using formal methods, which emphasizes the importance of focusing on the motivation of authors and updating simplistic formal approach to assessment.

\appendix

\section{Kruskal Wallis Test}\label{app:kruskel}
The hypothesis: the median values of journal metrics for Publication Concerns, Active PPJ, Inactive PPJ groups were equal.
We used the Kruskal-Wallis criterion \cite{kruskal1952a}. There were statistically significant differences in the journal metrics depending on the journal group, see Tables \ref{tab:stat_krus} and \ref{tab:rank_krus}.

\begin{table}[h!]
\caption{Statistics: Kruskal Wallis Test with grouping variable Journals}
\label{tab:stat_krus}
\centering
\begin{tabular}{|l|l|l|}
\hline
2016 & SNIP & SJR \bigstrut\\
\hline
Kruskal-Wallis H & 30.401 & 32.206    \bigstrut\\
\hline
Degrees of Freedom & 2 & 2         \bigstrut\\
\hline
Asymptotic Significance & 0.000 & 0.000     \bigstrut\\
\hline
\end{tabular}

\vspace{5mm}
\caption{Ranks: Kruskal Wallis Test}
\label{tab:rank_krus}
\centering
\begin{tabular}{|l|l|l|l|}
\hline
2016           & Journals             & Count   & Mean Rank \bigstrut\\
\hline
\multirow{4}{*}{SNIP} & Publication Concerns & 239 & 288.98    \bigstrut\\
\cline{2-4}
           & Active PPJ           & 193 & 279.30    \bigstrut\\
\cline{2-4}
           & Inactive PPJ         & 101 & 191.50    \bigstrut\\
\cline{2-4}
           & Total                & 533 &           \bigstrut\\
\hline
\multirow{4}{*}{SJR}  & Publication Concerns & 239 & 263.55    \bigstrut\\
\cline{2-4}
           & Active PPJ           & 194 & 307.32    \bigstrut\\
\cline{2-4}
           & Inactive PPJ         & 101 & 200.37    \bigstrut\\
\cline{2-4}
           & Total                & 534 &           \bigstrut\\
\hline
\end{tabular}
\end{table} 

\section{Rules for Awarding Academic Degrees} \label{app:kaz_new_rules}
According to the Rules\footnote{First introduced in the order of the Minister of Education and Science of the Republic of Kazakhstan dated March 31, 2011, No. 127} for awarding academic degrees of the Ministry of Education and Science of the Republic of Kazakhstan, the dissertation is written under the guidance of domestic and foreign supervisors who have academic degrees and are specialists in the field of scientific research of doctoral students. The main findings of the dissertation are to be published in at least 7 publications on the topic of the dissertation, including at least 3 in scientific publications recommended by the authorized body, 1 in an international scientific publication that has a non-zero impact factor in Web of Science or is indexed in Scopus, 3 in the materials of international conferences, including 1 in the materials of foreign conferences\footnote{See  \url{http://web.archive.org/save/http://adilet.zan.kz/rus/archive/docs/V1100006951/31.03.2011} and \url{https://academy-gp.kz/?page_id=71&lang=en}}.


\section*{Acknowledgments}
The authors would like to thank Dmitrii Marin (University of Waterloo, Canada) and Alexei Lutay (Russian Foundation for Basic Research, Russia) for helpful detail feedback and stimulating discussions. 

\begin{table}
    \centering
    \caption{List of Potentially Predatory Journals} \label{tab:list_ppj}
    \textit{ (the table has been removed for legal concerns) }    
\end{table}

\bibliographystyle{spbasic}


%








\end{document}